\documentclass{aa}  

\usepackage{threeparttable}
\usepackage{comment}

\usepackage{graphicx}
\usepackage{txfonts}
\usepackage{lipsum}

\usepackage{lscape}             
                                
\usepackage{placeins}

\newcommand{\ixpe}{\emph{IXPE}}
\newcommand{\source}{1E 2259$+$586}
\newcommand{\xmm}{\emph{XMM-Newton}}

\begin{document}

   \title{X-ray polarimetry of a magnetar in the aftermath of an anti-glitch}

   \subtitle{Revisiting 1E 2259$+$586 with \ixpe}

   \author{Ruth M. E. Kelly\inst{1}\corrauth{ruth.kelly@unipd.it}        
        \and Roberto Taverna\inst{1}\email{roberto.taverna@unipd.it}
        \and Roberto Turolla\inst{1,2}\email{roberto.turolla@unipd.it}
        \and Lorenzo Marra\inst{3}\email{lorenzo.marra@inaf.it}
        \and Swati Ravi\inst{4}\email{swatir@mit.edu}
        \and Jeremy Heyl\inst{5}\email{heyl@phas.ubc.ca}
        \and Silvia Zane\inst{2}\email{s.zane@ucl.ac.uk}
        \and Mingyu Ge\inst{6}\email{gemy@ihep.ac.cn}
        \and Hanlong Peng\inst{7}\email{penghl@ihep.ac.cn}
        \and Hua Feng\inst{6}\email{hfeng@ihep.ac.cn}
        \and Sandro Mereghetti\inst{8}\email{sandro.mereghetti@inaf.it}
        \and Herman Marshall\inst{4}\email{hermanm@mit.edu}
        \and Alice Borghese\inst{9}\email{alice.borghese@gmail.com}
        \and Michela Rigoselli\inst{8}\email{michela.rigoselli@inaf.it}
        \and Shuang-Nan Zhang\inst{6}\email{zhangsn@ihep.ac.cn}
        \and Gian Luca Israel\inst{10}\email{gianluca.israel@inaf.it}
        }

   \institute{Universit\`a di Padova, Dipartimento di Fisica e Astronomia, via Marzolo 8, I-35131 Padova, Italy
   \and Mullard Space Science Laboratory, University College London, Holmbury St Mary, Dorking, Surrey RH5 6NT
   \and INAF - Istituto di Astrofisica e Planetologia Spaziali, Via del Fosso del Cavaliere 100, I-00133, Roma
   \and MIT Kavli Institute for Astrophysics and Space Research, Massachusetts Institute of Technology, 77 Massachusetts Avenue, Cambridge, MA 02139, USA
   \and University of British Columbia, Department of Physics and Astronomy, Vancouver, BC V6T 1Z4, Canada
   \and State Key Laboratory of Particle Astrophysics, Institute of High Energy Physics, Chinese Academy of Sciences, Beijing 100049, China
   \and School of Physics and Technology, Nanjing Normal University, Nanjing, 210023, Jiangsu, China
   \and INAF, Istituto di Astrofisica Spaziale e Fisica Cosmica di Milano, via Corti 12, I-20133 Milano, Italy
   \and European Space Agency (ESA), European Space Astronomy Centre (ESAC), Camino Bajo del Castillo s/n, 28692 Villanueva de la Ca\~{n}ada, Madrid, Spain
   \and INAF - Osservatorio Astronomico di Roma, via Frascati 33, I-00078 Monteporzio Catone, Italy
    }

   \date{Received September 30, 20XX}

  \abstract
   {Magnetars have been known to display timing anomalies such as glitches and anti-glitches. In particular, the anomalous X-ray pulsar \source, whose properties in the X-ray band were recently investigated through a joint \xmm/\ixpe\ observation campaign, has been observed to have multiple glitches and three anti-glitches, the most recent occurring in May 2025. Although the X-ray emission is currently decaying, the flux, which roughly doubled at the onset of the anti-glitch, had not yet returned to its original level at the time of the \ixpe\ observation, and the pulse profile has also shown significant evolution.}
   {We present an analysis of the recent \ixpe\ observation of \source\ that spans $\sim 11$ days, from late April to early May 2026,  after its anti-glitch event in 2025. We compare the post-anti-glitch polarisation results with those obtained from the previous \ixpe\ observation in 2023. This provides an unprecedented opportunity to monitor the spectro-polarimetric evolution of a magnetar in the wake of a hallmark event such as an anti-glitch.} 
   {A dedicated $400\,\mathrm{ks}$ \ixpe\ Director’s Discretionary Time (DDT) observation was approved to track possible changes in the degree and angle of polarisation with respect to the pre-anti-glitch campaign.
   Spectral and polarimetric properties were jointly modelled within the {\sc xspec} and {\sc ixpeobssim} suites, and the analysis was further expanded using a maximum-likelihood technique.}
   {No significant detection was found in the phase-averaged and energy-integrated spectrum of this observation. However, despite the significantly shorter exposure, polarisation was clearly detected in the phase-resolved spectrum and the degree of polarisation was found to have a higher best-fitting value, though still statistically consistent, than the previous measurement from 2023, reaching a level $\gtrsim30\%$ in correspondence to the pulse profile minimum. In spite of some slight differences between the two observations, the polarisation trends remain largely consistent and indicate that the source has mostly returned to its pre-event state.}
   {}


   \maketitle
   \nolinenumbers

\section{Introduction}
X-ray polarimetry offers a unique opportunity to study complex physical phenomena otherwise inaccessible to us. In particular, the polarimetric observations of highly magnetised, isolated neutron stars could prove key to finally probing the complex magnetic field structures expected around them and the behaviour of matter in extreme magnetic environments. Significantly, X-ray polarimetry may be crucial to finally testing vacuum birefringence, a quantum-electrodynamics (QED) effect that makes magnetised vacuum birefringent  \cite[][]{heisenberg_folgerungen_1936}.

Magnetars, with typical dipole spin-down magnetic fields of $B\sim10^{14}$--$10^{15} \ \mathrm{G}$ are extremely interesting candidates for X-ray polarimetric observations \citep{turolla_magnetars_2015}. Powered by their magnetic energy, these isolated neutron stars  are characterised by spin periods $P\approx 1$--$12\,\mathrm s$, spin-down rates $\dot{P} \approx 10^{-15}$--$10^{-10}\,\mathrm{s\,s^{-1}}$, and X-ray luminosities $L_{\mathrm{X}} \approx 10^{30}$--$10^{36}\,\mathrm{erg\,s^{-1}}$. At soft X-ray energies ($0.5$--$10 \ \mathrm{keV}$), magnetar spectra are typically well-fitted by a two-component model, either two blackbodies or a blackbody plus a power-law \cite[][]{mereghetti_strongest_2008}. The non-thermal component is generally thought to arise from resonant Compton scattering (RCS) of photons onto (mildly relativistic) particles streaming along the closed field lines of the twisted magnetar magnetosphere \cite[][]{thompson_electrodynamics_2002}, while the thermal components may originate from limited regions of the surface, either covered by a gaseous atmosphere or with the condensed surface exposed \cite[see e.g.][and references therein]{taverna_x-ray_2020}.

In the highly magnetised environment of a magnetar, radiation propagates in two orthogonal polarisation modes, with the electric field oscillating either parallel or perpendicular to the plane of the magnetic field and the propagation direction for ordinary (O) and extraordinary (X) modes, respectively.
Models have shown that emission from a highly magnetised atmosphere is expected to produce high degrees of polarisation ($\sim 80 \%$) predominantly in the X-mode, while emission from condensed surface is characterised by a lower polarisation fraction ($\sim20\%$), either in the O- or X-mode. Magnetospheric reprocessing by RCS should instead result in about $\sim 33 \%$ polarisation in the X-mode \citep{fernandez_x-ray_2011,taverna_x-ray_2020}. 

In December 2021, NASA-ASI collaboration launched the Imaging X-ray Polarimetry Explorer \cite[\ixpe,][]{weisskopf_imaging_2022} to observe polarised X-rays in the $2$--$8 \ \mathrm{keV}$ range, opening up a new window into understanding the physics of highly magnetised neutron stars. To date, \ixpe\ has observed six magnetars: 4U 0142+61 \citep{taverna_polarized_2022}, 1RXS J170849.0--400910 \cite[][]{zane_strong_2023}, SGR 1806--20 \citep{turolla_ixpe_2023}, 1E 2259+586 \citep{heyl_detection_2024}, 1E 1841--045 \citep{rigoselli_ixpe_2025, stewart_x-ray_2025} and 1E 1547.0--5408 \citep{stewart_vacuum_2026, taverna_long_2026}. However, owing to the intrinsically photon-hungry nature of polarimetric observations \cite[which demand significantly larger photon statistics than conventional spectral and timing techniques,][]{bellazzini_x-ray_2010}, combined with the limitations imposed by the slow spacecraft repointing, 
some of the most dynamic manifestations of magnetar activity, such as the emission of bursts/flares and the occurrence of glitches \cite[][]{mereghetti_strongest_2008, turolla_magnetars_2015}, remain essentially unexplored from a polarimetric perspective.

In the midst of the development of future facilities with better temporal responsiveness, the most viable approach to investigate fast transient events with current instrumentation is to target those accompanied by long-lasting afterglows and to monitor the evolution of the polarimetric properties before and after their onset. Within the sample of magnetars targeted by \ixpe\ so far, the anomalous X-ray pulsar (AXP) \source\ stands out as one of the most promising candidates for such an investigation. Located at the centre of CTB 109 supernova remnant (G109.1-1.0), this source exhibits a spin period $P \approx 7 \ \mathrm{s}$ and a period derivative $\dot{P} \approx 0.05 \times 10^{-11} \ \mathrm{s\,s^{-1}}$ \cite[][]{dib_16_2014}, for an inferred dipolar magnetic field $\approx6\times10^{13}\,\mathrm{G}$. Although the spin-down rate is small, compared to other members of the class, the source has experienced two previous anti-glitch events, i.e. an abrupt decrease in spin frequency, in April 2012 \cite[][]{archibald_anti-glitch_2013} and April 2019 \cite[][]{younes_radiatively_2020}. The former, characterised by a fractional frequency variation $\Delta\nu/\nu \approx -3\times 10^{-7}$, was accompanied by  the presence of bursting activity, a substantial enhancement of the persistent X-ray flux, and a moderate change in the pulse profile.

\ixpe\ originally observed the source in 2023, for a total exposure time of $1.2 \ \mathrm{Ms}$ \cite[][]{heyl_detection_2024}. Data were processed along with a simultaneous observation from \xmm\ \cite[][]{jansen_xmm-newton_2001} and partially simultaneous observations by the Neutron star Interior Composition Explorer \cite[NICER;][]{gendreau_neutron_2016}. The analysis revealed a relatively low energy- and phase-integrated polarisation degree ($\mathrm{PD}=5.6 \pm 1.4 \%$), with a polarisation angle $\mathrm{PA} = -75.2^\circ \pm 7.4^\circ$. A significant detection was obtained only in the range of $2$ -- $3.2 \ \mathrm{keV}$ ($\mathrm{PD}=6.1 \pm 1.5 \%$, $\mathrm{PA} = -66.4^\circ \pm 7.1^\circ$), while a phase-dependent analysis showed a variable $\mathrm{PD}$ ranging from $\sim0$ to $\sim25 \%$ \cite[][]{heyl_detection_2024}.

This behaviour aligns with the model proposed by \citet{pizzocaro_detailed_2019}, in which the observed radiation originates from two hot spots on the stellar surface, one of them being threaded by a magnetic loop populated by charged particles that resonantly scatter thermal photons via RCS. Indeed, previous spectral observations revealed a soft X-ray spectrum well fitted by the usual blackbody$+$power-law decomposition, but with the addition of a phase-dependent absorption feature centred at $\approx 1 \ \mathrm{keV}$, attributed to resonant scattering onto protons. Within this framework, the surface magnetic field would be locally much stronger than inferred from spin-down estimates, reaching values up to $10^{14}$--$10^{15} \ \mathrm{G}$ \citep{pizzocaro_detailed_2019, heyl_detection_2024, taverna_x-ray_2024}. This could provide a natural justification for the occurrence of the observed anti-glitches, in terms of the emergence of a localised magnetic field structures beneath, or associated with, the hot spots of the surface \citep{lyutikov_magnetospheric_2013}. Furthermore, the relatively low polarisation degree measured by \ixpe\ has been interpreted as evidence for magnetic condensed surface emission \cite[][]{brinkmann_thermal_1980,turolla_bare_2004,potekhin_radiative_2012,taverna_x-ray_2020}, while photons scattered within the magnetic loop are expected to exhibit polarisation degrees close to the $\approx30\%$ limit, as for RCS onto mildly relativistic particles.

In May 2025, \source\ underwent a new, third, anti-glitch event around MJD 60820 detected by NICER and the Einstein Probe telescope \cite[EP;][]{peng_large_2026}. This anti-glitch was reminiscent of the one observed in April 2012 \citep{archibald_anti-glitch_2013}. As in 2012, the event was accompanied by a short burst detected by the Fermi Gamma-ray Burst Monitor \citep{meegan_fermi_2009} around MJD 60817 \citep{godwin_17_2026}, and the persistent X-ray flux roughly doubled with respect to its pre-event level \citep{peng_large_2026}. 
This enhancement was accompanied by a marked change in the pulse profile, particularly around the two minima, with a flatter plateau appearing between the two peaks \cite[H.-L. Peng private communication; see also][]{peng_large_2026}. In the following months, the source displayed a gradual decline in flux (although it had not yet returned to its quiescent level), together with a progressive evolution of the pulse profile back towards its pre-event shape. These changes prompted a DDT observation with \ixpe, which was carried out at the end of April 2026 (P.I. Lorenzo Marra) for a total exposure time of $\approx 400\,\mathrm{ks}$. At the time of the 2026 \ixpe\ observation, the flux remained enhanced by $\sim13\%$, and the pulse profile had almost, but not fully recovered \citep{peng_large_2026}.

In this paper, we present an analysis of this recent \ixpe\ observation of \source\ after the anti-glitch event in May 2025. In section \ref{observation} we detail the observation and data analysis procedure. The timing and spectro-polarimetric results are presented in section \ref{results}. Discussion and conclusions follow in sections \ref{discussion} and \ref{conclusions}.

\section{Observation and data analysis} \label{observation}
\source\ ($\mathrm{R.A.}=345^\circ.2839$, $\mathrm{DEC}=58^\circ.879$) was re-observed by \ixpe\ starting on 24 April, 2026 at 07:32:21.050 UTC following the detection by NICER and EP of an anti-glitch around MJD 60820 \cite[25 May 2025;][]{peng_large_2026}. The observation had a total exposure time of $\sim 400 \ \mathrm{ks}$, ending on 5 May 2026 at 14:51:46.564 UTC. The level 1 and level 2 photon lists were downloaded from the \ixpe\ online archive\footnote{https://heasarc.gsfc.nasa.gov/docs/ixpe/archive/}.
The photon arrival times were corrected to the Solar System barycentre using the \textsc{ftool} \textsc{barycorr} with Development Ephemeris DE421 from the publicly available \textsc{HEASoft}\footnote{https://heasarc.gsfc.nasa.gov/docs/software/lheasoft/download.html} suite. Background rejection was applied following the procedure described in \citet[][see also \citealt{di_marco_hitchhikers_2026}]{di_marco_handling_2023}\footnote{https://github.com/aledimarco/IXPE-background}, upgraded to take into account the patch deployed to fix the recent failure experienced by the detector unit (DU) 2. 
The source and background regions were extracted with the procedure described in \cite{piconcelli_evidence_2004}, using the SAOImage Deep Space 9 software \cite[\textsc{DS9};][]{joye_new_2003} to maximise the signal-to-noise ratio in all three DUs. This resulted in circular source regions with radii $r^{\mathrm{src}} = 75'', 65'', 65 ''$ for $\mathrm{DU1, DU2, DU3}$, respectively, and annular background regions with inner radii $r^{\mathrm{bkg}}_{\mathrm{in}} = 90''$ and outer radii $r^{\mathrm{bkg}}_{\mathrm{out}} = 210''$.
Barycentered and background-subtracted level 2 files were finally processed using the most up-to-date response functions (version 20260424) publicly available\footnote{https://heasarc.gsfc.nasa.gov/docs/ixpe/caldb/} in the HEASARC \ixpe\ calibration database {\sc caldb} and a weighted analysis was performed within {\sc xspec} \citep{arnaud_xspec_1996} and using the most recent version of the {\sc ixpeobssim} software \cite[][]{baldini_ixpeobssim_2022}\footnote{Response files version 20260701 were used for the \textsc{ixpeobssim} analysis, the most up-to-date version compatible with this type of analysis.}.

\begin{figure}
    \centering
    \includegraphics[width=\linewidth]{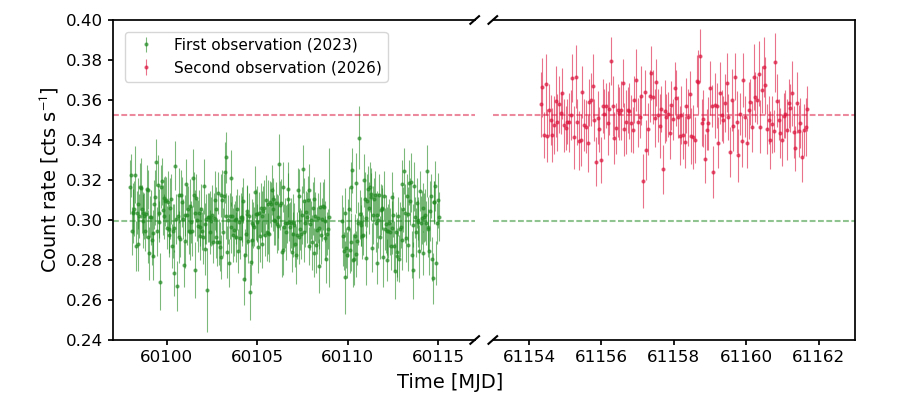}
    \caption{Background-subtracted light curves of \source\ in the $2$--$8 \ \mathrm{keV}$ energy band for the 2023 and 2026 \ixpe\ observations, obtained by summing the three DUs and binned in $4 \ \mathrm{ks}$ intervals. Error bars are $1\sigma$ statistical uncertainties, and the dashed lines indicate the mean count rate of each observation.}
    \label{fig:lightcurves}
\end{figure}

Figure \ref{fig:lightcurves} shows the background-subtracted light curves in the $2$--$8 \ \mathrm{keV}$ energy band for the 2023 and 2026 \ixpe\ observations, extracted using the same source and background regions and summing the three DUs. As can be seen, the mean count rate for the 2026 observation is $\sim17\%$ higher than in 2023, further indicating that the source had not yet completely returned to its pre-event flux level at the time of the second observation with \ixpe.

\section{Results} \label{results}
\subsection{Timing solution} \label{sec:timing}
We searched for a timing solution using the \textsc{stingray/hendrics} package\footnote{https://hendrics.stingray.science/en/latest/} \citep{bachetti_hendrics_2018}. The most prominent peak in the power spectrum was found at $\sim0.29\,\mathrm{Hz}$. Given the known double-peaked morphology of the source pulse profile \citep{fahlman_x-ray_1981, pizzocaro_detailed_2019, heyl_detection_2024}, this value is attributed to the first harmonic of the fundamental frequency, which, after performing a $Z_n^2$ search, we identified as $\nu = 0.143280(1)\,\mathrm{Hz}$. We then calculated the event times of arrival with \textsc{HENphaseogram} (splitting the observation into $128$ bins) and fitted them with a linear spin-down model using \textsc{PINT}\footnote{https://github.com/nanograv/PINT} \citep{luo_pint_2021, susobhanan_pint_2024}. The best-fitting frequency turned out to be
$\nu=0.14328020\pm1.5\times10^{-7}\,\mathrm{Hz}$ (error at $68\%$ c.l.) at epoch $61158.02492384862\,\mathrm{MJD}$, corresponding to a spin period, $P=6.9793314\pm7.1\times10^{-6} \ \mathrm{s}$, in agreement within $1\sigma$ with values previously reported in the literature \cite[see][]{olausen_mcgill_2014,heyl_detection_2024}.
Owing to the shorter exposure time of the present \ixpe\ observation, the spin period derivative could not be fully constrained. The pulsed fraction, calculated as $\mathrm{PF_{P2P}} = ({N_{\mathrm{max}} - N_{\mathrm{min}})}/({N_{\mathrm{max}} + N_{\mathrm{min}}})$ resulted in $\mathrm{PF_{P2P}}=33.53 \pm 0.16 \%$, consistent with the $\sim 35 \%$ pulse fraction reported in \cite{heyl_detection_2024}. Finally, we used the \textsc{xpphase} tool of \textsc{ixpeobssim} to assign a rotational phase to each event using this timing solution. Phase zero corresponds to the start of the observation (MJD 61154.31899, Barycentric Dynamical Time (TDB), referred to the Solar System barycentre) and is therefore not tied to a specific feature of the pulse profile.

\subsection{Phase-averaged polarimetric analysis: \textsc{ixpeobssim} and \textsc{QUEEN-BEE}} \label{sec:phase-ave-ixpeobssim}
An initial polarisation analysis was performed with the \textsc{ixpeobssim} package\footnote{https://ixpeobssim.readthedocs.io/en/latest/overview.html} version 33.0.0 \citep{baldini_ixpeobssim_2022}, using the \textsc{xpselect}, \textsc{xpbin} and \textsc{pcube} tools. In the phase- and energy-integrated data polarisation turned out to be below the Minimum Detectable Polarisation at $99\%$ confidence level \citep[$\mathrm{MDP}_{99}$;][]{weisskopf_understanding_2010}. A significant signal was detected only in the $3$--$4 \ \mathrm{keV}$ range, with $\mathrm{PD} = 11.1 \pm 3.3 \%$ and $\mathrm{PA} = 80.1^\circ \pm 18.4^\circ$. Interestingly, when dividing the energy into $0.5 \ \mathrm{keV}$ intervals, a significant detection is found only in the $3$--$3.5 \ \mathrm{keV}$ band ($\mathrm{PD} = 15.0 \pm 3.9 \%$ and $\mathrm{PA} = 86.0^\circ \pm 7.5^\circ$).\footnote{Unlike the observation of 2023 \citep{heyl_detection_2024}, no significant detection was found in the range $2$--$3.2 \ \mathrm{keV}$.} 

An additional model-independent polarimetric analysis was performed with the event-based Bayesian framework \textsc{QUEEN-BEE} (version 3.3) using its constant polarisation model \texttt{scout\_const} \citep{ravi_queen-bee_2025, ravi_whats_2026}.
QUEEN-BEE analyses smooth polarisation variability, probing both linear time- and energy-dependent PA rotation and PD evolution across the observation. We compare the time-rotating PA model \texttt{scout\_rot}, the energy-rotating PA model \texttt{scout\_energy\_rot}, and the energy-evolving PD model \texttt{scout\_energy\_evol} against both a constant-polarisation model and the unpolarised null hypothesis following the prescriptions outlined in \cite{ravi_whats_2026} and Ravi et al. (in prep.).

The QUEEN-BEE analysis strongly favours the unpolarised model for the phase-averaged data, yielding $\Delta \ln Z \approx 5.7$ when compared to a constant-polarisation model\footnote{$\ln Z$ is the logarithmic evidence i.e. the natural log of a Bayesian evidence. $\Delta\ln Z$ quantifies the relative Bayesian evidence between two models, interpreted through the modified Jeffreys scale \citep{jeffreys_theory_1939}}. This does not imply an absence of intrinsically polarised emission, since phase-dependent variations in PA can lead to substantial cancellation in the phase-averaged Stokes parameters. The inferred Stokes parameters under the constant-polarisation hypothesis are consistent with negligible polarisation, with $q=-0.028^{+0.017}_{-0.019}$ and $u=0.001^{+0.022}_{-0.020}$ ($90\%$ credible interval). More complex models allowing for temporal or spectral polarisation evolution are likewise not favoured. While a model allowing polarisation angle to vary linearly with energy is formally the highest-evidence polarised model, it exceeds the unpolarised model by only $\Delta\ln Z \approx 1.9$, below the cut-off for decisive evidence according to the modified Jeffreys scale \citep{jeffreys_theory_1939} and has an inferred rotation rate which is highly unconstrained. Models  permitting a linear rotation of PA with time and PD with energy yield negative $\Delta \ln Z$ values compared to the unpolarised model.

However, as in the \textsc{ixpeobssim} analysis, in the $3.0$--$4.0\, \mathrm{keV}$ energy bin, QUEEN-BEE does find evidence for polarised emission. The constant-polarisation model is preferred over the unpolarised hypothesis, with $\Delta\ln Z \approx 3.2$. The inferred Stokes parameters are $q=-0.098^{+0.045}_{-0.040}$ and $u=0.046^{+0.043}_{-0.036}$ ($90\%$ credible interval), corresponding to  $\mathrm{PD} = 10.8 \pm 4.5\% $ and $\mathrm{PA} = 77.5 \pm 11.5^\circ$. Models allowing PA to vary linearly with time, and PD to evolve linearly with energy, as well as the unpolarised model are disfavoured relative to the constant-polarisation hypothesis.
Interestingly, a model that allows PA evolution with energy yields the highest evidence among the tested models, with an improvement of $\Delta\ln Z \approx 2.2$ relative to the constant-polarisation model and decisive evidence of polarisation compared to the unpolarised hypothesis $\Delta\ln Z \approx 5.45$. The evidence is therefore decisive for polarised emission in this band, but only modest for an energy-dependent PA relative to constant polarisation. The inferred energy-dependent rotation rate is $\omega_E=-55.8^{+31.1}_{-32.2}~{\rm deg~keV^{-1}}$, suggesting a possible decrease in PA with increasing energy.

\subsection{Phase-averaged spectro-polarimetric analysis: \textsc{xspec}} \label{sec:phase-ave-xspec}

Following the initial polarisation analysis, a complementary spectro-polarimetric analysis was performed using \textsc{xspec} \cite[version 12.15.1;][]{arnaud_xspec_1996}. Here we restrict our analysis to the $2$--$6 \ \mathrm{keV}$ range, above which the background dominates, and bin the data such that each bin has a signal-to-noise ratio $\gtrsim 3$. Interstellar absorption was taken into account using the \textsc{tbabs} model, with abundances from \cite{wilms_absorption_2000}, and a normalisation factor was introduced to ensure cross-calibration between all three detectors.

\begin{figure*}
    \centering
    \includegraphics[width=0.45\linewidth]{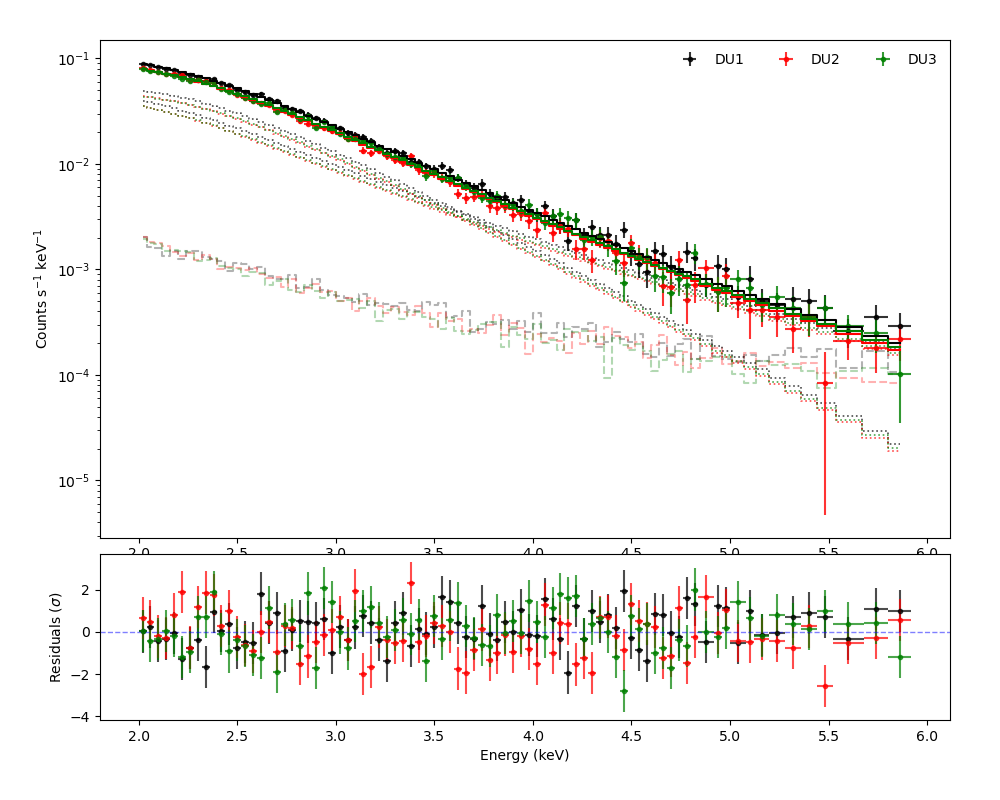}
    \includegraphics[width=0.45\linewidth]{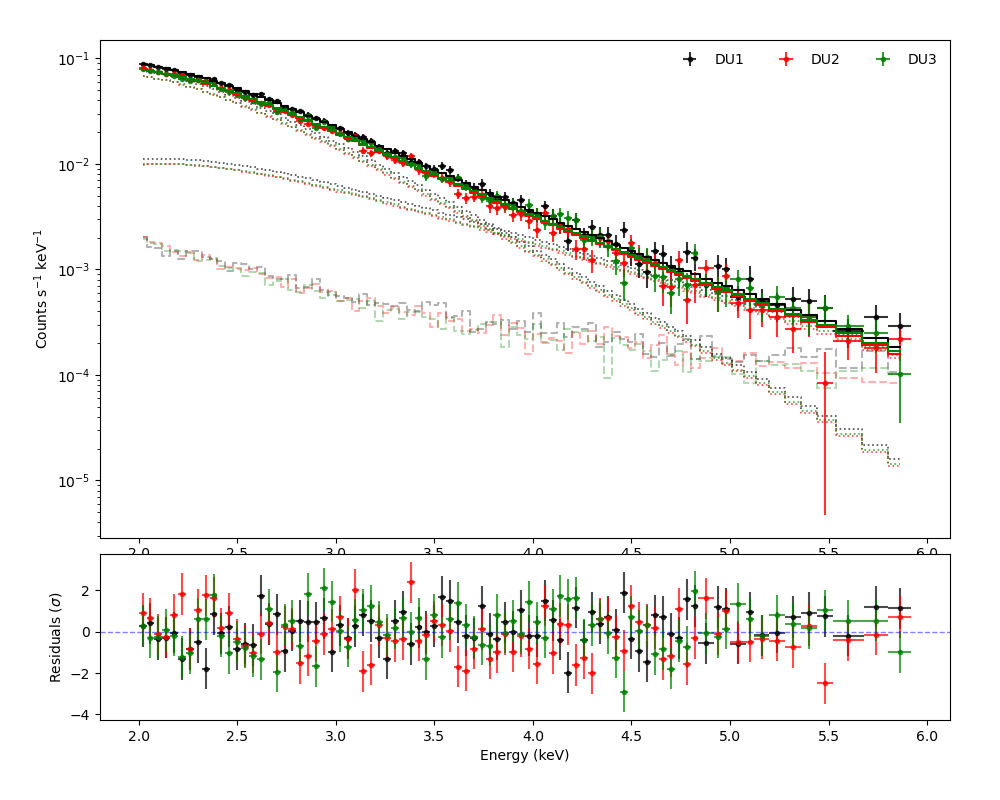}
    \caption{Phase-averaged count spectra for the data collected by \ixpe\ detectors DU1 (black), 2 (red) and 3 (green) fit by the BB+PL model (left) and BB+BB model (right; see Table \ref{tab:phase_ave_spectral_fit}). The residuals of the fit are shown in the lower panel.}
    \label{fig:spec}
\end{figure*}

We find that two-component models (two blackbodies or blackbody plus power-law) give acceptable fits, while a single blackbody component is rejected. In models that leave $N_{H}$ as a free parameter, the value is unconstrained due to the limited energy range of \ixpe\ not extending below $2 \ \mathrm{keV}$; we therefore fix it to the values obtained from the EP spectral fits for each decomposition, $N_\mathrm{H} = 0.96 \times 10^{22}\,\mathrm{cm}^{-2}$ for the blackbody+power-law model and $0.64 \times 10^{22}\,\mathrm{cm}^{-2}$ for the double blackbody model \cite[H.-L. Peng, priv. comm.; see also][]{peng_large_2026}.
Both the blackbody+power-law (BB+PL; \textsc{const}\,$\times$\,\textsc{tbabs}\,$\times$\ (\textsc{bbodyrad+powerlaw})) and blackbody+blackbody (BB+BB; \textsc{const}\,$\times$\,\textsc{tbabs}\,$\times$\,(\textsc{bbodyrad+bbodyrad})) decompositions provide statistically acceptable fits, with $\chi^2/\mathrm{dof} = 223.36/246$ and $225.36/246$, respectively (see Figure~\ref{fig:spec} and Table~\ref{tab:phase_ave_spectral_fit}). Since both two-component models provide acceptable fits, we adopt the BB+PL decomposition for consistency with \cite{heyl_detection_2024}.

An alternative model including a Gaussian absorption line was also fit to the data (BB+PL+Line; \textsc{const} $\times$ \textsc{phabs} $\times$ (\textsc{bbodyrad+powerlaw}) $\times$ \textsc{gabs}), freezing the absorption constant and line parameters to the values reported in \cite{heyl_detection_2024}. The model statistics are as acceptable as the BB+PL model and the blackbody and power-law parameters agree across both fits. As the absorption line falls below the \ixpe\ range, we opt to use the BB+PL model for the remainder of our analysis.

\begin{table}[]
    \centering
    \renewcommand{\arraystretch}{2}
    \setlength{\tabcolsep}{8pt}
    \caption{Phase-averaged spectral fits.}
    \label{tab:phase_ave_spectral_fit}
    \begin{tabular}{c c c}
        \hline\hline
         & BB+PL & BB+BB \\
        \hline
        $N_\mathrm{H}$ ($10^{22}\,\mathrm{cm}^{-2}$) & $0.96$\tablefootmark{(a)} & $0.64$\tablefootmark{(a)} \\
        $kT_\mathrm{BB,1}$ (keV) & $0.444^{+0.015}_{-0.012}$ & $0.407^{+0.013}_{-0.015}$ \\
        $R_\mathrm{BB,1}$\tablefootmark{(b)} (km) & $2.91^{+0.36}_{-0.34}$ & $4.30^{+0.30}_{-0.24}$ \\
        $kT_\mathrm{BB,2}$ (keV) & $-$ & $0.83^{+0.11}_{-0.08}$ \\
        $R_\mathrm{BB,2}$\tablefootmark{(b)} (km) & $-$ & $0.43^{+0.17}_{-0.13}$ \\
        $\Gamma_\mathrm{PL}$ & $3.70^{+0.20}_{-0.23}$ & $-$ \\
        $\mathrm{norm_{PL}}$\tablefootmark{(c)} & $40.0^{+12.1}_{-11.5}$ & $-$ \\
        constant$_\mathrm{DU1}$ & $1$\tablefootmark{(d)} & $1$\tablefootmark{(d)} \\
        constant$_\mathrm{DU2}$ & $0.970 \pm 0.009$ & $0.970 \pm 0.009$ \\
        constant$_\mathrm{DU3}$ & $0.974 \pm 0.009$ & $0.974 \pm 0.009$ \\
        $F_\mathrm{unabs}$\tablefootmark{(e)} & $2.08 \pm 0.01$ & $1.99 \pm 0.01$ \\
        $F_\mathrm{obs}$\tablefootmark{(e)} & $1.82 \pm 0.01$ & $1.82 \pm 0.01$ \\
        $\chi^2$/dof & $223.36/246$ & $225.36/246$ \\
        \hline\hline
    \end{tabular}
    \tablefoot{
        Fits for the data in the $2$--$6\,\mathrm{keV}$ range, using the \textsc{const}\,$\times$\,\textsc{tbabs}\,$\times$\,(\textsc{bbodyrad+powerlaw}) (BB+PL) and \textsc{const}\,$\times$\,\textsc{tbabs}\,$\times$\,(\textsc{bbodyrad+bbodyrad}) (BB+BB) decompositions.
        \tablefoottext{a}{Frozen to the values obtained from the EP spectral fits for each decomposition \cite[H.-L. Peng, priv. comm.; see also][]{peng_large_2026}.}
        \tablefoottext{b}{Calculated from \textsc{xspec} $\mathrm{norm_{BB}}$ assuming a distance of $3.2\,\mathrm{kpc}$ \citep{heyl_detection_2024, pizzocaro_detailed_2019}.}
        \tablefoottext{c}{Units of $10^{-3}\,\mathrm{photons\,cm^{-2}\,s^{-1}\,keV^{-1}}$ at $1\,\mathrm{keV}$.}
        \tablefoottext{d}{Frozen parameter.}
        \tablefoottext{e}{Fluxes in units of $10^{-11}\,\mathrm{erg\,cm^{-2}\,s^{-1}}$ in the $2$--$6\,\mathrm{keV}$ band.}
    }
\end{table}

The polarisation measurements were extracted from \textsc{xspec} by simultaneously fitting the Stokes $I$, $Q$ and $U$ spectra, following the procedure detailed in \citet[][see also \citealt{taverna_long_2026}]{strohmayer_x-ray_2017}. We freeze the model parameters to those obtained for the best-fitting BB+PL spectral decomposition and include a constant polarisation model (\textsc{polconst}).
Here again, we do not find a significant polarisation detection in the phase-average, energy-integrated spectrum, while a significant detection of $\mathrm{PD} = 11.9 \pm 3.6 \%$ ($\mathrm{PA} = 85.6^\circ \pm 8.9^\circ$) is found in the $3$--$4 \ \mathrm{keV}$ bin, in agreement with the model-independent results obtained using \textsc{ixpeobssim} and \textsc{QUEEN-BEE}.

\subsection{Phase-resolved spectro-polarimetric analysis}
In order to investigate the phase-resolved behaviour, we split the pulse profile into seven equal bins using \textsc{xpselect}, following the timing analysis and solution described in Sect.~\ref{sec:timing}. As before, we perform an initial polarisation analysis using \textsc{ixpeobssim}, and a complementary analysis using \textsc{xspec}, extracting the source and background regions in each phase bin. We limit our analysis to the energy-integrated band of $2$--$6 \ \mathrm{keV}$ due to the background dominating above $6 \ \mathrm{keV}$.

\begin{figure}
    \centering
    \includegraphics[width=\linewidth]{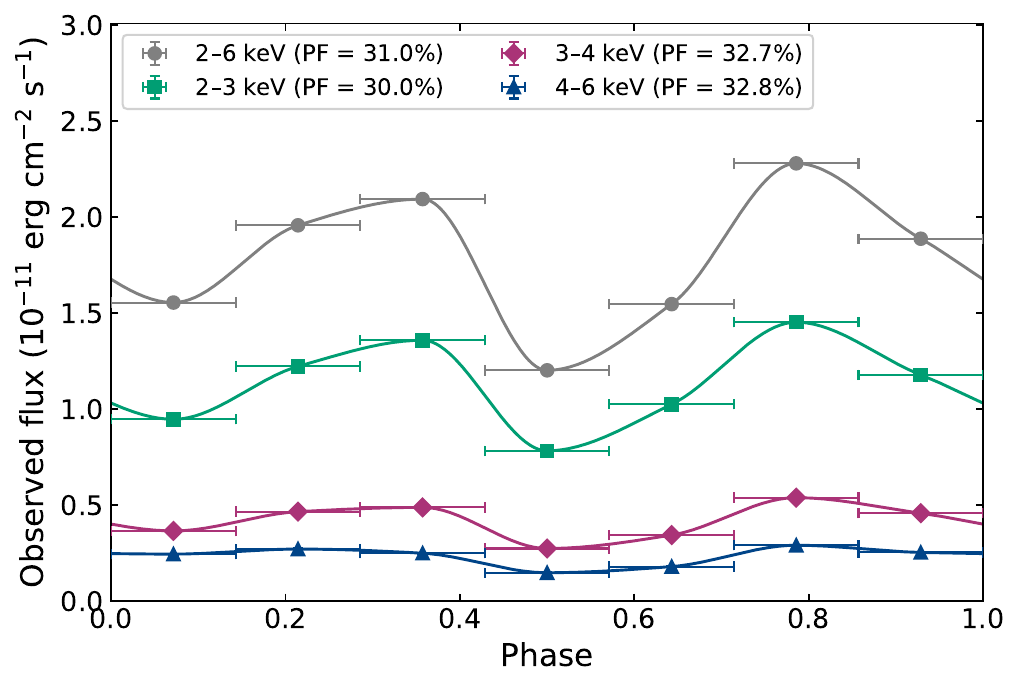}
    \caption{Pulse profile in the $2$--$3 \ \mathrm{keV}$ (green squares), $3$--$4 \ \mathrm{keV}$ (pink-purple diamonds), $4$--$6 \ \mathrm{keV}$ (blue triangles) and $2$--$6 \ \mathrm{keV}$ (grey circles) bins. The flux and pulsed fraction ($\mathrm{PF_{P2P}}$) are calculated from \textsc{xspec} using the best-fit BB+PL model. The curves are interpolated between points for illustrative purposes. The errors in flux are smaller than the marker size and therefore not visible in the plot.}
    \label{fig:pulse_energy}
\end{figure}

Figure \ref{fig:pulse_energy} shows the energy-resolved pulse profile calculated in \textsc{xspec} using the best-fit (BB+PL) model. The pulse profile is broadly similar across the different energy bands, while the $2$--$3 \ \mathrm{keV}$ band contributes the largest flux and the pulsed fraction is slightly higher above $3 \ \mathrm{keV}$. The pulsed fraction values in Figure \ref{fig:pulse_energy} are computed from the model fluxes in the seven phase bins, while the fraction in Sect.~\ref{sec:timing} uses counts in the finer bins.

\begin{table*}[]
    \centering
    \renewcommand{\arraystretch}{2}
    \setlength{\tabcolsep}{10pt}
    \caption{Phase-resolved spectral fit.}
    \label{tab:phase_res_spectral_fit}
    \begin{tabular}{c c c c c c c c}
        \hline\hline
        Phase bin & $kT_\mathrm{BB}$ (keV) & $R_\mathrm{BB}$\tablefootmark{(a)} $\ (\mathrm{km})$ & $\Gamma$ & $\mathrm{norm_{PL}}$\tablefootmark{(b)} & ${F_\mathrm{unabs}}$\tablefootmark{(c)} & ${F_\mathrm{obs}}$\tablefootmark{(c)} & $\chi^2$/dof \\
        \hline
        $0.00$--$0.14$ & $0.416^{+0.023}_{-0.013}$ & $3.53^{+0.43}_{-0.81}$ & $2.71^{+0.78}_{-1.25}$ & $9.6^{+22.6}_{-8.4}$ & $1.773^{+0.019}_{-0.020}$ & $1.555^{+0.017}_{-0.017}$ & $138.64/198$ \\
        $0.14$--$0.29$ & $0.431^{+0.036}_{-0.016}$ & $3.45^{+0.82}_{-0.83}$ & $3.30^{+0.58}_{-1.09}$ & $25.3^{+32.0}_{-20.9}$ & $2.238^{+0.022}_{-0.022}$ & $1.957^{+0.019}_{-0.019}$ & $169.87/177$ \\
        $0.29$--$0.43$ & $0.429^{+0.033}_{-0.016}$ & $3.82^{+0.89}_{-0.83}$ & $3.50^{+0.64}_{-1.38}$ & $29.4^{+38.6}_{-26.1}$ & $2.404^{+0.022}_{-0.022}$ & $2.094^{+0.019}_{-0.019}$ & $151.83/168$ \\
        $0.43$--$0.57$\tablefootmark{(d)} & $0.414^{+0.026}_{-0.012}$ & $3.44^{+0.40}_{-0.93}$ & $2.82^{+1.25}_{-2.99}$ & $6.0^{+30.7}_{-5.9}$ & $1.379^{+0.017}_{-0.017}$ & $1.202^{+0.015}_{-0.015}$ & $148.58/186$ \\
        $0.57$--$0.71$ & $0.405^{+0.021}_{-0.011}$ & $4.15^{+0.46}_{-0.84}$ & $3.01^{+0.89}_{-1.58}$ & $10.2^{+28.4}_{-9.5}$ & $1.779^{+0.019}_{-0.019}$ & $1.547^{+0.016}_{-0.017}$ & $176.44/192$ \\
        $0.71$--$0.86$ & $0.450^{+0.046}_{-0.033}$ & $2.95^{+1.25}_{-0.68}$ & $3.96^{+0.42}_{-0.50}$ & $75.2^{+40.8}_{-37.8}$ & $2.614^{+0.023}_{-0.023}$ & $2.280^{+0.020}_{-0.020}$ & $165.83/195$ \\
        $0.86$--$1.00$ & $0.456^{+0.050}_{-0.031}$ & $2.70^{+1.06}_{-0.65}$ & $3.73^{+0.53}_{-0.66}$ & $44.0^{+33.3}_{-27.3}$ & $2.158^{+0.021}_{-0.021}$ & $1.888^{+0.018}_{-0.019}$ & $168.32/198$ \\
        \hline\hline
    \end{tabular}
    \tablefoot{
        Fits for the data in the $2$--$6 \ \mathrm{keV}$ range, using the $\textsc{const} \times \textsc{tbabs} \times (\textsc{bbodyrad+powerlaw})$ spectral model with $N_{H}$ frozen at $0.96\times10^{22}\,\mathrm{cm}^{-2}$.
        \tablefoottext{a}{Calculated from \textsc{xspec} $\mathrm{norm_{BB}}$ assuming a distance of $3.2 \ \mathrm{kpc}$ \citep{heyl_detection_2024, pizzocaro_detailed_2019}.}
        \tablefoottext{b}{Units of $10^{-3} \ \mathrm{photons\,cm^{-2}\,s^{-1}\,keV^{-1}}$ at $1\,\mathrm{keV}$.}
        \tablefoottext{c}{Fluxes in units of $10^{-11}\,\mathrm{erg\,cm^{-2}\,s^{-1}}$ in the $2$--$6\,\mathrm{keV}$ band.}
        \tablefoottext{d}{The power-law is poorly constrained in this phase bin.}
    }
\end{table*}

\begin{figure}
    \centering
    \includegraphics[width=\linewidth]{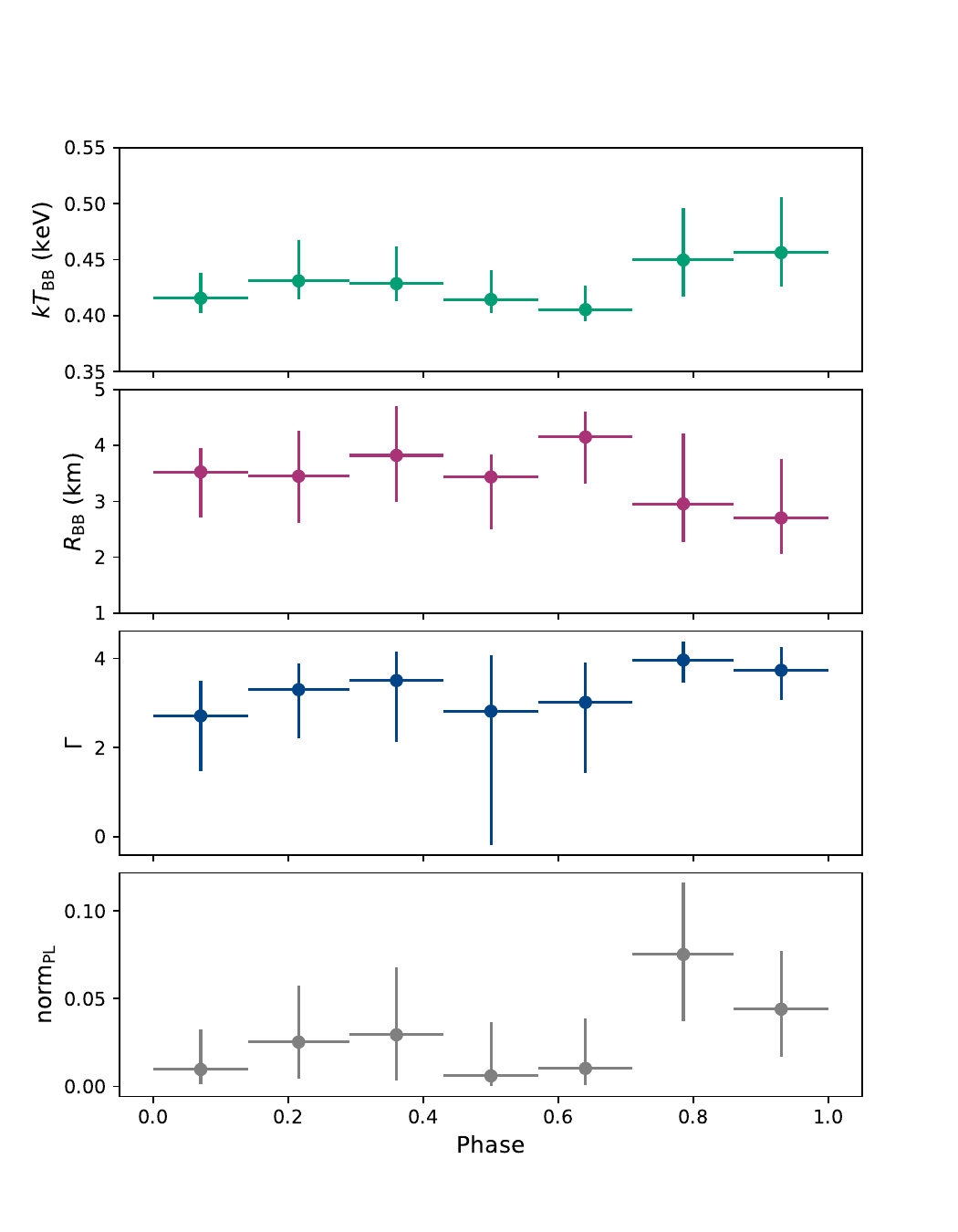}
    \caption{Temperature (panel 1) and radius (panel 2) of the blackbody, and photon index (panel 3) and normalisation (panel 4) of the power-law spectral components as a function of the rotation phase for the model reported in Table \ref{tab:phase_res_spectral_fit}.}
    \label{fig:spec_params}
\end{figure}

In the case of the \textsc{xspec} analysis, we fit the BB+PL model to the data, leaving all spectral parameters free in each phase bin, and report the best-fit parameters for each phase bin in Table \ref{tab:phase_res_spectral_fit}. The evolution of the spectral components with rotational phase is shown in Figure \ref{fig:spec_params}.  The blackbody component is well constrained across all phases of rotation, with a temperature that remains relatively consistent at $\sim0.41$--$0.46 \ \mathrm{keV}$. $R_\mathrm{BB}$ ranges from $\sim 2.7$ to $\sim 4.2 \ \mathrm{km}$ in size, without any significant variation across the phases, remaining small, consistent with emission from a small hot spot. We find that the power-law component is well constrained only in the 0.71–0.86 and 0.86–1.00 phase bins, while in the other bins its parameters carry large uncertainties. Its normalisation is lowest in the $0.00$--$0.14$, $0.43$--$0.57$ and $0.57$--$0.71$ bins, and consistent with zero in the flux minimum ($0.43$--$0.57$), suggesting that the spectrum could be dominated by the thermal component in these phases, which may reflect the changing viewing angle of the hot spot as the star rotates. The spectral values are broadly consistent with those reported in \cite{heyl_detection_2024}, although the power-law component is somewhat flatter in the present observation ($\Gamma\approx3.7$, compared with $\Gamma\approx4.4$), a difference that may be partly due to the different absorption column, energy range and inclusion of the absorption line in their model.

Once again to probe the polarisation degree and angle, we fit the data in \textsc{xspec} with the BB+PL model, freezing the parameters to those of Table \ref{tab:phase_res_spectral_fit}, and add an additional \textsc{polconst} parameter, which we leave free.
The best-fit phase-resolved polarisation results are detailed in Table \ref{tab:phase_res_polarisation}, where all values are obtained from the same fit and reported with $1\sigma$ uncertainties. The marked rows indicate the phase bins in which polarisation is detected at $\geq 3\sigma$ (closed $3\sigma$ contour in the $q$–$u$ plane), while the other rows give the best-fitting values in bins without a significant detection.
The \textsc{ixpeobssim} analysis for the phase-dependent results follows the method detailed in Sect.~\ref{sec:phase-ave-ixpeobssim}, this time running the \textsc{xpbin} and \textsc{pcube} tools for each phase bin. 

\begin{figure}
    \centering
    \includegraphics[width=\linewidth]{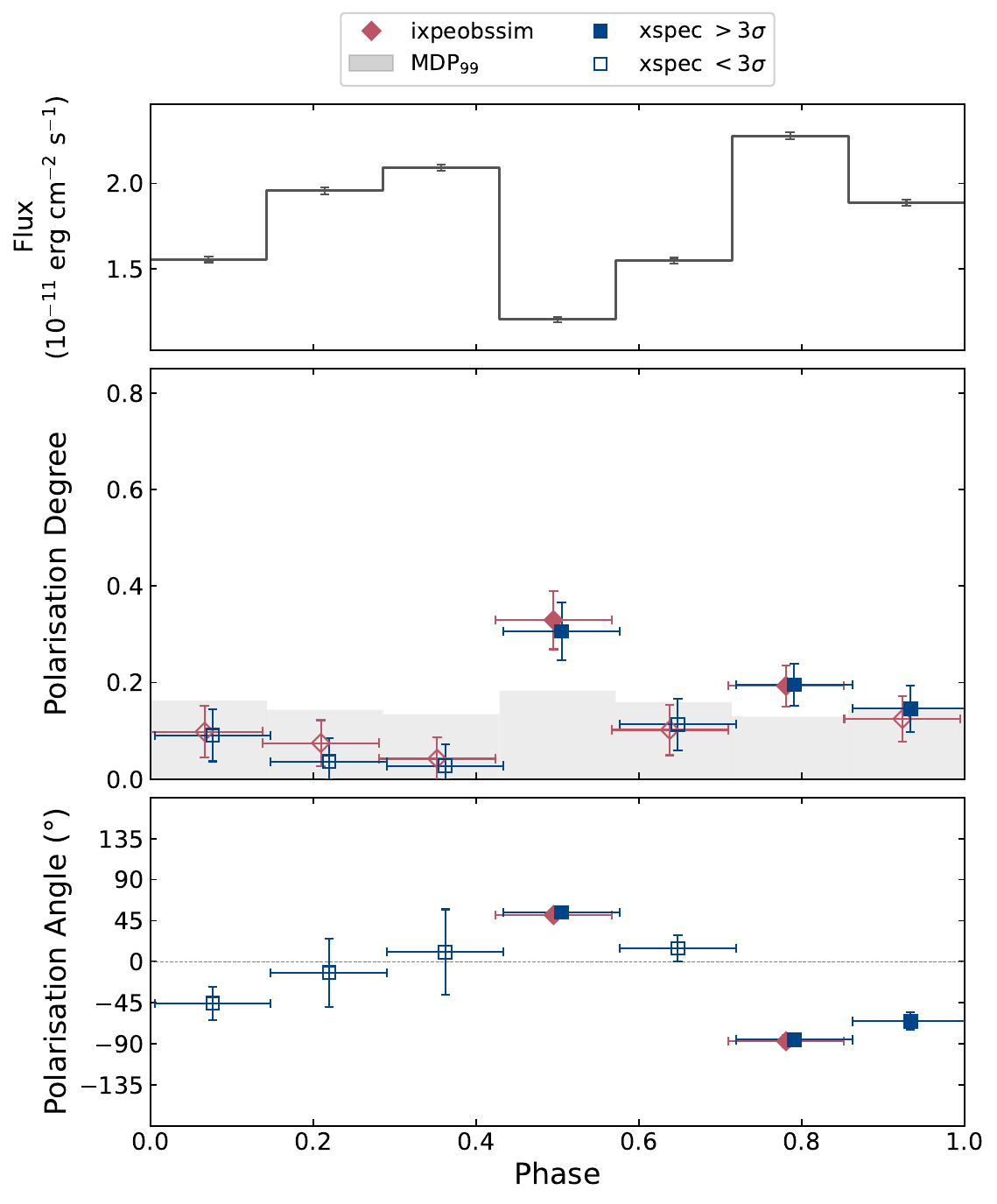}
    \caption{Phase-dependent polarisation degree (middle) and angle (bottom) produced using \textsc{ixpeobssim} and \textsc{xspec}, where the former is shown by pink-purple diamonds and the latter is shown by blue squares. Both show energy-integrated quantities over the $2$--$6 \ \mathrm{keV}$ range. Filled markers indicate a closed $3\sigma$ contour for \textsc{xspec} and a significant detection above $\mathrm{MDP}_{99}$ for \textsc{ixpeobssim}, while empty markers indicate the best-fitting values for non-significant detections. The observed flux is shown in the top panel for reference. All errors are shown at $1\sigma$ confidence.}
    \label{fig:phase_pol}
\end{figure}

Phase-resolved polarisation measurements from both \textsc{ixpeobssim} and \textsc{xspec} are shown in Figure \ref{fig:phase_pol}. The \textsc{xspec} results are shown as blue points where the filled squares indicate a significant ($3\sigma$) detection and the empty ones the best-fitting value in case of a non-significant detection. The results obtained with \textsc{ixpeobssim} are indicated by pink-purple diamonds. The MDP obtained with \textsc{ixpeobssim} in each phase is indicated by the grey bars, and the observed flux is shown in grey for comparison. Using the convention from \cite{heyl_detection_2024}, the $0.43$--$0.57$ bin corresponds to the `big dip', $0.71$--$0.86$ to the `big peak', $0.86$--$1.00$ and $0.00$--$0.14$ to the `little dip', and $0.14$--$0.43$ to the `little peak'.

The analysis with both methods finds significant detections in only some of the phase bins, with the highest detection observed near the absolute minimum of the X-ray flux (the `big dip', phase $0.43$--$0.57$) where the thermal (BB) component dominates the emission. Here, the polarisation degree is $\mathrm{PD} \gtrsim 30 \%$ and the corresponding $\mathrm{PA} \approx 55^\circ$.
There appears to be a $\sim 40 ^\circ$  shift in the value of the polarisation angle, measured near the absolute minimum of the flux (phase $0.43$--$0.57$) and the `big peak' (phase $0.71$--$0.86$), where the polarisation degree drops to $\sim 20 \%$. This shift in polarisation angle increases to $\sim 60 ^\circ$ difference between phase bins $0.43$--$0.57$ and $0.86$--$1.0$. No significant polarisation was detected in any of the other phases using these methods.

\begin{table}[]
    \centering
    \renewcommand{\arraystretch}{2}
    \caption{Phase-resolved polarisation results.}
    \label{tab:phase_res_polarisation}
    \begin{tabular}{c c c c}
        \hline\hline
        Phase bin & PD (\%) & PA ($^\circ$) & $\chi^2$/dof \\
        \hline
        $0.00$--$0.14$ & $9.1 \pm 5.4$ & $-45.5 \pm 18.1$ & $243.73/316$ \\
        $0.14$--$0.29$ & $3.6^{+4.8}_{-3.6}$ & $-11.9 \pm 37.5$\tablefootmark{(b)} & $286.38/295$ \\
        $0.29$--$0.43$ & $2.7^{+4.5}_{-2.7}$ & $10.6 \pm 46.8$\tablefootmark{(b)} & $267.63/286$ \\
        $0.43$--${0.57}$\tablefootmark{(a)} & ${30.6 \pm 6.0}$ & $54.0 \pm 5.7$ & ${244.40/304}$ \\
        $0.57$--$0.71$ & $11.3 \pm 5.3$ & $14.7^{+14.0}_{-14.1}$ & $291.60/310$ \\
        ${0.71}$--${0.86}$\tablefootmark{(a)} & ${19.6 \pm 4.3}$ & ${-85.5 \pm 6.4}$ & ${264.26/313}$ \\
        ${0.86}$--${1.00}$\tablefootmark{(a)} & ${14.6 \pm 4.8}$ & ${-65.4^{+9.7}_{-9.5}}$ & ${305.69/316}$ \\
        \hline\hline
    \end{tabular}
    \tablefoot{
        Results from \textsc{xspec} analysis for the data in the $2$--$6 \ \mathrm{keV}$ range using the model parameters in Table \ref{tab:phase_res_spectral_fit}.
        All rows are obtained from the same fit. All values are reported with $1\sigma$ uncertainties, calculated with the \textsc{err} function in \textsc{xspec} unless marked otherwise. \tablefoottext{a}{Phase bins in which the $3\sigma$ confidence contour in the $q$--$u$ plane is closed, i.e. polarisation is detected at $\geq 3\sigma$; in the other bins the values are best-fitting values without a significant detection.}
        \tablefoottext{b}{The \textsc{err} function did not converge for the polarisation angle in this phase bin, as the lower bound of the polarisation degree reaches zero; the uncertainty quoted is the $1\sigma$ standard error from the fit.}
}
\end{table}

To investigate the energy dependence of the phase-resolved polarisation, we repeated the analysis in the $2$--$3$, $3$--$4$ and $4$--$6 \ \mathrm{keV}$ bands using the same phase binning. No significant detections were found above $4 \ \mathrm{keV}$, while only one significant detection was found in the $3$--$4 \ \mathrm{keV}$ band ($\mathrm{PD} = 32.7 \pm 8.4 \%$ and  $\mathrm{PA} = 87.5^\circ \pm 7.6^\circ$ from \textsc{xspec}) in the $0.71$--$0.86$ phase bin corresponding to the `big peak', where the power-law component is largest, in line with the resonantly scattered emission interpretation suggested in \cite{heyl_detection_2024}. In the $2$--$3 \ \mathrm{keV}$ band, this same phase bin is also significantly polarised ($\mathrm{PD} = 19.7 \pm 5.9 \%$ and $\mathrm{PA} = -82.1^\circ \pm 8.8^\circ$). Additional significant detections are found in the $2$--$3 \ \mathrm{keV}$ band in the same phases as in the $2$--$6 \ \mathrm{keV}$ band ($\mathrm{PD} = 38.9 \pm 8.1 \%$, $\mathrm{PA} = 45.0^\circ \pm 6.1^\circ$ in $0.43$--$0.57$ and $\mathrm{PD} = 24.3 \pm 6.6 \%$, $\mathrm{PA} = -69.5^\circ \pm 7.9^\circ$ in $0.86$--$1.00$). The \textsc{ixpeobssim} results agree with the \textsc{xspec} results within $1 \sigma$.

Finally, for completeness, we also applied the \textsc{QUEEN-BEE} analysis described in Sect.~\ref{sec:phase-ave-ixpeobssim} to each phase bin in the $2$--$6 \ \mathrm{keV}$ band, comparing the constant-polarisation model, energy-evolving PD model \texttt{scout\_energy\_evol} and the time- and energy-rotating PA models \texttt{scout\_rot} and \texttt{scout\_energy\_rot} against the unpolarised null hypothesis. Polarisation models are decisively preferred over the un-polarised model in the $0.43$--$0.57$, $0.71$--$0.86$ and $0.86$--$1.00$ phase bins, in agreement with the \textsc{xspec} and \textsc{ixpeobssim} results. In the $0.43$--$0.57$ and $0.71$--$0.86$ bins, a constant-polarisation model is favoured over the models with time- and energy-dependent polarisation, while in the $0.86$--$1.00$ an energy-rotating PA model is decisively preferred over both the unpolarised ($\Delta \ln{Z} \approx 7.8$) and constant-polarisation ($\Delta \ln{Z} \approx 6.6$) models, with $\omega_E = 38.6^{+23.3}_{-15.1} \ \mathrm{deg} \, \mathrm{keV^{-1}}$  ($90\%$ credible interval). This preference appears to be driven largely by the $4$--$6 \ \mathrm{keV}$ photons, where the polarisation is not individually detected, as the PAs measured separately in the $2$--$3$ and $3$--$4 \ \mathrm{keV}$ bands differ by less than this rate implies. Again, in agreement with \textsc{xspec} and \textsc{ixpeobssim}, no polarised model is significantly favoured over the unpolarised hypothesis in the remaining phase bins.

\section{Discussion} \label{discussion}
The enhancement of flux that accompanied the anti-glitch event in May 2025, alongside the modification to the pulse profile immediately following the event, prompted the observation by \ixpe\ to test whether the event had been accompanied by a reconfiguration of the magnetosphere or a change in the surface physical state, such as a potential phase transition from a condensed surface (assumed to explain the previous \ixpe\ observation) to an atmosphere. A comparison of the polarisation properties observed in 2026 and in 2023, while the source was in quiescence, is therefore key.

The  polarisation measured by \ixpe\ with this most recent observation of \source\ is very low, and while 
there are some differences in the phase-averaged polarisation observed during the two observations, the differences can be attributed to the different statistical quality of the data. Namely, in the 2023 observation there was a significant detection \cite[$\mathrm{PD}=5.6\pm1.4 \%$;][]{heyl_detection_2024} observed in the energy-integrated spectrum, while no such detection was found in the 2026 observation. This lack of detection does not necessarily mean a physical change to the source and could simply be due to the lower observation time ($\sim 400 \ \mathrm{ks}$ in comparison to $\sim 1.2 \ \mathrm{Ms}$), resulting in the polarisation falling below the higher $\mathrm{MDP_{99}}$ threshold (found to be $\sim 6 \%$ from the \textsc{ixpeobssim} analysis), and the already low polarisation, caused by the partial cancellation of the Stokes parameters as the polarisation angle rotates with phase. The 2026 non-detection is therefore statistically compatible with the polarisation measured in 2023. As discussed below, the phase-resolved polarisation remains largely consistent between the two observations and it is only the phase-averaged signal that falls below the $\mathrm{MDP_{99}}$ for this shorter exposure time.
Additionally, the 2023 observation found $\mathrm{PD}=6.1\pm1.5 \%$ in the $2$--$3.2\ \mathrm{keV}$ range, while the 2026 observation shows a detection in the slightly higher energy range above $3 \ \mathrm{keV}$ ($\mathrm{PD} = 11.9 \pm 3.6 \%$ and $\mathrm{PD} = 15.0 \pm 3.9 \%$ in the $3$--$4\ \mathrm{keV}$ and $3$--$3.5\ \mathrm{keV}$ energy ranges, respectively).
This apparent shift can also be explained by the different sensitivities rather than a physical change, with the lack of a significant detection in the $2$--$3.2\ \mathrm{keV}$ band in 2026 resulting from the $\mathrm{MDP_{99}}(\approx6.4\%$ from the \textsc{ixpeobssim} analysis) exceeding 2023 detected PD, while the significant detection above $3\ \mathrm{keV}$ remains compatible with the upper-limit of $14.6 \%$ detected by \cite{heyl_detection_2024} for $3.2$--$5\ \mathrm{keV}$.

\begin{figure}
    \centering
    \includegraphics[width=\linewidth]{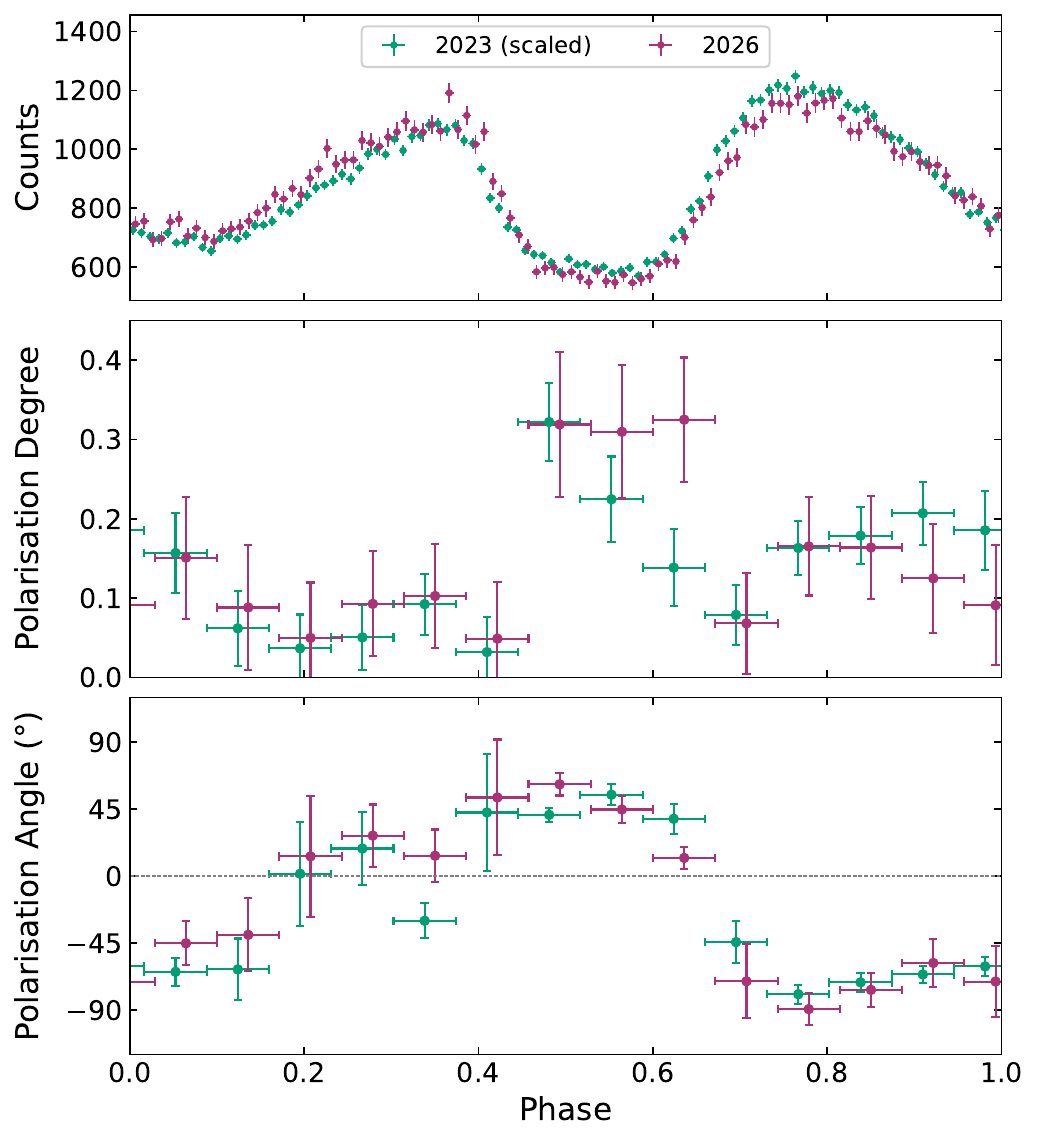}
    \caption{Comparison between the energy-integrated, phase-resolved polarisation observed from \source\ by \ixpe\ in 2023 (green) and 2026 (purple). The results were produced using the likelihood method described in \citet[][see also \citealt{heyl_detection_2024} for details on the application with the 2023 observation]{gonzalez-caniulef_unbinned_2023}. Both observations are shown in the phase convention adopted in this paper, and polarisation angles are shown in the range $-90^\circ$ to $+90^\circ$, as in Figure \ref{fig:phase_pol}.
    }
    \label{fig:2023/2026}
\end{figure}

Focusing on the phase-dependent spectrum and polarisation properties, we find only a small change in the pulse profile and polarisation properties between the two observations (see Figure \ref{fig:2023/2026}). We used the same likelihood method as applied in the analysis of the 2023 observation, fitting the Stokes parameters directly for each photon \cite[][see also \citealt{gonzalez-caniulef_unbinned_2023} for details on the method]{heyl_detection_2024}. The counts from the 2023 observation have been scaled by the ratio of the total source counts in 2026 to those in 2023 for display to allow for a direct comparison. We find that the pulse profile, polarisation degree and angle remain relatively consistent across both observations, indicating that the source has mostly returned to its pre-anti-glitch state almost a year after the event occurred. However, a noticeable difference between the two observations appears in the absolute minimum of the pulse profile \cite[referred to as the `big dip' in ][]{heyl_detection_2024}. Here we can see that the flux is increased with respect to the earlier observation. Additionally, while this minimum remains at  the phase of  the strongest polarisation in both observations, the best-fitting polarisation degree is higher in the 2026 observation ($\sim30\%$ compared to $\sim22\%$ obtained by combining the 2023 `big dip' phase bins of \cite{heyl_detection_2024}, a difference of $\sim1.3 \sigma$), remaining $\sim30 \%$ across the entire minimum, instead of decreasing across the minimum as in the 2023 observation (Figure \ref{fig:2023/2026}).

The emission observed in 2023 was explained by \cite{heyl_detection_2024} to originate from a condensed surface hot spot covered by a local plasma loop. In this scenario, the polarisation degree in the minimum, `big dip', is explained by the moderately polarised O-mode photons originating at the condensed surface passing through the loop, while the lower polarisation in the `big peak' and `little dip' comes from the photons scattered preferentially into X-mode when the loop does not intersect the line-of-sight but is still in view.
We see a $\sim40$--$60^\circ$ rotation, consistent across both observations, reproduced by \cite{heyl_detection_2024} for the 2023 observation using a rotating vector model and including a mode-switch. The net overall polarisation angle swing is less than  $90 ^\circ$ because the intrinsic mode-switch combines with the different emission contributions and the viewing geometry. We note, however, that \cite{heyl_detection_2024} also found an acceptable (but slightly less preferred) fit for a different rotating vector model without the mode-switch. A similar mixing of emission components could explain the energy-dependent polarisation angle rotation in the $0.86$--$1.00$ phase bin suggested by the \textsc{QUEEN-BEE} analysis, with thermal photons leaving the hot spot in the O-mode dominating at lower energies, while at higher energies those resonantly scattered in the loop, and hence converted to the X-mode, dominate.
It should also be noted that, in previous observations of the source, the emission has been modelled by two antipodal hot spots, one of which is covered by a plasma loop \citep{pizzocaro_detailed_2019}.

Considering these results together, the 2026 polarisation and pulse morphology are broadly similar to those observed in 2023. Within the condensed-surface-plus-plasma-loop scenario of \cite{heyl_detection_2024}, the anti-glitch does not appear to have resulted in a permanent change to the emission geometry of the source. Since the polarisation in the most recent observation is in good agreement across most of the pulse phases, it suggests that the large-scale surface and magnetospheric configuration is recovering towards its pre-event (quiescent) state. However, almost a year after the event, the recovery was not yet fully complete: at the time of the second  \ixpe\ observation, the flux was still $\sim13 \%$ above its pre-event state, and the pulse profile has not yet fully recovered its previous shape \citep{peng_large_2026}. 
Additionally, the local magnetic field loop of higher $B$ seems to still be present with a consistent phase and polarisation angle, and could suggest that it is a manifestation of a stable multi-polar component. 

The only feature not accounted for in this picture is the increased flux, together with the higher, though not significantly so, polarisation degree observed in the pulse minimum, which may indicate that there has been a residual change to the surface emission. As briefly mentioned, one possibility we aimed to test was whether this anti-glitch had been accompanied by a phase transition of the outermost layer of the neutron star from condensed surface to atmosphere, causing the change in pulse profile immediately after the event. Although such a phase transition seems unlikely due to the higher polarisation expected in such a scenario ($\sim 80 \%$), we cannot completely rule it out. As this observation took place almost a year after the anti-glitch event, it is possible that there was a phase-transition immediately after the event that then reverted to a condensate as the flux decayed. However, this re-transition does not explain the still elevated polarisation degree in the flux minimum with respect to the previous 2023 observation and is not supported by any observational evidence. Perhaps instead this change in polarisation implies a modest change to the geometry of the plasma loop itself, while the bulk of the emission topology has relaxed back to its pre-event state. We emphasise that this is speculative, the data require only that the plasma loop persists, a change in geometry is simply a tentative suggestion to explain the elevation in flux and polarisation in the phase minimum.

\section{Conclusions} \label{conclusions}
In this paper, we analyse the recent re-observation of \source\ by \ixpe\ following an anti-glitch in May 2025. We do not detect significant polarisation in the phase-averaged and energy-integrated spectrum, a result statistically compatible with the 2023 measurement given the shorter exposure. However, we observe $\mathrm{PD} = 11.9 \pm 3.6 \%$ and $\mathrm{PA} = 85.6^\circ \pm 8.9^\circ$ in the phase-averaged $3$--$4\ \mathrm{keV}$ bin. 
The phase-resolved profile indicates a significant polarisation detection ($30.6 \pm 6.0 \%$ in the $2$--$6\ \mathrm{keV}$ band) in the phase corresponding to the flux minimum, an increase over, though still consistent with, the $\mathrm{PD} \sim22 \%$ detection from the previous 2023 observation \citep{heyl_detection_2024}, and reaching almost $40 \%$ when restricted to $2$--$3\ \mathrm{keV}$. The polarisation in the other phases remains lower ($\lesssim 20 \%$). 
There appears to be a difference of $\sim 60^\circ$ in the polarisation angle between the `big dip' and the other phases. Additionally, and not to be confused with the PA rotation in phase, the \textsc{QUEEN-BEE} analysis decisively favours polarised emission in the $3$--$4 \ \mathrm{keV}$ energy bin, with more modest evidence for a polarisation angle rotation ($\sim 55^\circ \ \mathrm{keV^{-1}}$) relative to constant polarisation, and, in the $0.86$--$1.00$ phase bin, favours a PA rotating with energy at $\sim 40^\circ \ \mathrm{keV^{-1}}$, although this result appears to be driven partly by the $4$--$6 \ \mathrm{keV}$ photons, where polarisation is not independently detected.
The polarisation results suggest that, at the time of observation, \source\ had almost completely returned to its pre-anti-glitch emission topology, with no evidence for a persistent large-scale change in its emission geometry at that epoch; however, these data cannot distinguish between the absence of a substantial change at the time of the event and a reconfiguration that subsequently relaxed.
Future observations of this source to monitor the evolution of the polarisation will be needed to confirm this statement.
This work presents the first X-ray polarimetric comparison of a magnetar before and after an anti-glitch, and it shows that polarimetry can establish whether the emission geometry persists through such an event even while the flux and pulse profile are still recovering. However, polarimetric measurements obtained within weeks of the event, while the flux enhancement and pulse-profile changes are at their peak, are essential to reveal whether (and how) the magnetic topology is transiently reconfigured.

\begin{acknowledgements}
This research was supported by the International Space Science Institute (ISSI) in Bern, through the International Team project 25-657 `Polarimetric Insights into Extreme Magnetism'. The work of R.M.E. K., R. Ta., R. Tu. and L. M. is partially funded through the PRIN grant 2022 - project 2022LWPEXW - `An X-ray view of compact objects in polarized light', European Union funding - Next Generation EU, Mission 4 Component 1, CUP C53D23001180006. S. R. acknowledges support from NASA for project 80NSSC26K1006 `Energy-Dependent Polarization Rotation: Testing General Relativistic Disk Models with IXPE' under which the energy-dependent QUEEN-BEE models used in this work were developed. M.Y. G. thanks the support from the National Natural Science Foundation of China (Grant No. 12673057) and the China's Space Origins Exploration Program. SM and MR acknowledge support from the INAF grants `COMEFAR' (O.F. 1.05.24.01.08) and `Toward Neutron Stars Unification' (O.F. 1.05.24.02.08). AB acknowledges support through the European Space Agency (ESA) research fellowship programme.
\end{acknowledgements}

\bibliographystyle{bibtex/aa}
\bibliography{references}

\end{document}